\documentclass[
submission,
twocolumns,
5p,
a4paper,
oneside,
12pt]{elsarticle}

\usepackage{hyperref}
\usepackage{graphicx}
\usepackage{longtable}
\usepackage{dcolumn}
\usepackage{amsmath,amssymb,amsfonts,bm} 
\usepackage{upgreek}  
\usepackage{color}  
\usepackage{lineno}
\usepackage{tabularx}

\journal{Physics Letters A}

\begin{document} 
\begin{frontmatter}

\title{First measurement of the temperature dependence of muon 
       transfer rate from muonic hydrogen atoms to oxygen}

\author[1]{E.~Mocchiutti\corref{cor1}}
\ead{Emiliano.Mocchiutti@ts.infn.it}
\author[2]{A.~Adamczak}
\author[3]{D.~Bakalov}
\author[4]{G.~Baldazzi}
\author[5,6]{R.~Benocci}
\author[5]{R.~Bertoni}
\author[5,7]{M.~Bonesini}
\author[1]{V.~Bonvicini}
\author[1]{H.~Cabrera~Morales}
\author[5]{F.~Chignoli}
\author[5,7]{M.~Clemenza}
\author[8,9]{L.~Colace}
\author[1,10]{M.~Danailov}
\author[3]{P.~Danev}
\author[11,12]{A.~de~Bari}
\author[12]{C.~De~Vecchi}
\author[8,13]{M.~De~Vincenzi}
\author[1,14]{E.~Furlanetto}
\author[4,15]{F.~Fuschino}
\author[1,16,17]{K.~S.~Gadedjisso-Tossou}
\author[1,18]{D.~Guffanti}
\author[19]{K.~Ishida}
\author[4,15]{C.~Labanti}
\author[5,6]{V.~Maggi}
\author[5]{R.~Mazza}
\author[11,12]{A.~Menegolli}
\author[4,15]{G.~Morgante}
\author[5,7]{M.~Nastasi}
\author[16]{J.~Niemela}
\author[1]{C.~Pizzolotto}
\author[20,21]{A.~Pullia}
\author[20,22]{R.~Ramponi}
\author[4]{L.~P.~Rignanese}
\author[12]{M.~Rossella}
\author[14]{N.~Rossi}
\author[3]{M.~Stoilov}
\author[1,16]{L.~Stoychev}
\author[8]{L.~Tortora}
\author[1]{E.~Vallazza}
\author[1]{G.~Zampa}
\author[1,14,19]{A.~Vacchi}

\cortext[cor1]{Corresponding author}

\address[1]{ Sezione INFN di Trieste, via A. Valerio 2, Trieste, Italy}
\address[2]{ Institute of Nuclear Physics, Polish Academy of Sciences, Radzikowskiego 152, PL31342 Krak\'{o}w, Poland}
\address[3]{ Institute for Nuclear Research and Nuclear Energy, Bulgarian Academy of Sciences, blvd.\ Tsarigradsko ch.~72, Sofia 1142, Bulgaria}
\address[4]{ Sezione INFN di Bologna, viale Berti Pichat 6/2, Bologna, Italy}
\address[5]{ Sezione INFN di Milano Bicocca, Piazza della Scienza 3, Milano, Italy}
\address[6]{ Dipartimento di Scienze dell'Ambiente e della Terra, Universit\`a di Milano Bicocca, Piazza della Scienza 1, Milano, Italy}
\address[7]{ Dipartimento di Fisica G. Occhialini, Universit\`a di Milano Bicocca, Piazza della Scienza 3, Milano, Italy}
\address[8]{ Sezione INFN di Roma Tre, Via della Vasca Navale 84, Roma, Italy}
\address[9]{ Dipartimento di Ingegneria, Universit\`a degli Studi Roma Tre, Via V. Volterra 62, Roma, Italy} 
\address[10]{ Sincrotrone Elettra Trieste, SS14, km 163.5, Basovizza, Italy}
\address[11]{ Dipartimento di Fisica, Universit\`a di Pavia, via A.~Bassi 6, Pavia, Italy}
\address[12]{ Sezione INFN di Pavia, Via A.~Bassi 6, Pavia, Italy}
\address[13]{ Dipartimento di Matematica e Fisica, Universit\`a di Roma Tre, Via della Vasca Navale 84, Roma, Italy}
\address[14]{ Dipartimento di Scienze Matematiche, Informatiche e Fisiche, Universit\`a di Udine, via delle Scienze 206, Udine, Italy}
\address[15]{ INAF-OAS Bologna, via P.~Gobetti 93/3, Bologna, Italy}
\address[16]{ The Abdus Salam International Centre for Theoretical Physics, Strada Costiera 11, Trieste, Italy}
\address[17]{ Laboratoire de Physique des Composants \`a Semi-conducteurs (LPCS), D\'epartment de physique, Universit\'e de Lom\'e, Lom\'e, Togo}
\address[18]{ Gran Sasso Science Institute, via F. Crispi 7, L'Aquila, Italy}
\address[19]{ Riken Nishina Center, RIKEN, 2-1 Hirosawa, Wako, Saitama 351-0198, Japan}
\address[20]{ Sezione INFN di Milano, via Celoria 16, Milano, Italy}
\address[21]{ Dipartimento di Fisica, Universit\`a degli Studi di Milano, via Celoria 16, Milano, Italy}
\address[22]{ IFN-CNR, Dipartimento di Fisica, Politecnico di Milano, piazza Leonardo da Vinci 32, Milano, Italy}

\begin{abstract}
We report the first measurement of the temperature dependence of
muon transfer rate from muonic hydrogen atoms to oxygen between 100 and
300~K. Data were obtained from the X-ray spectra of delayed events
in a gaseous target, made of a H$_2$/O$_2$ mixture, exposed to a muon beam. This work
sets constraints on theoretical models of muon transfer and is of
fundamental importance for the measurement of the hyperfine
splitting of muonic hydrogen ground state as proposed by the FAMU collaboration.
\end{abstract}

\begin{highlights}
\item Temperature dependence of the $\mu$ transfer rate from $\mu$p to oxygen is measured
\item A strong monotonic rise by a factor three of the rate is observed between 104--300 K
\item This effect will be exploited to measure the hyperfine splitting of the $\mu$p 1$S$ state
\item Experiment took place at the RIKEN muon facility, Rutherford Appleton Laboratory
\item Fast scintillator counters  based on LaBr3 crystals read by photomultipliers are used
\end{highlights}

\begin{keyword}
Muonic hydrogen \sep Transfer rate \sep Oxygen
\PACS 25.39.Mr \sep 34.70.+e \sep 36.10.Ee
\end{keyword}

\end{frontmatter}


\section{Introduction}
  We present the results of a systematic experimental investigation of the
  temperature dependence of the muon transfer rate from the ground-state muonic
  hydrogen atom $\mu$p to oxygen. A precise knowledge of this process
  is of key importance for the final objective of the FAMU
  Collaboration --
  the measurement of the hyperfine splitting (HFS) of the 1$S$ state of
  $\mu$p by means of laser spectroscopy~\cite{bakalov15a,adamczak16}.

  In the experiment planned by the FAMU collaboration, a laser tuned to the
  hyperfine-splitting resonance energy induces singlet-to-triplet
  transitions between the total-spin states $F=0$ and $F=1$ of 1$S$
  muonic hydrogen, which is formed and then thermalized in a gaseous
  H$_2$ target. After the laser excitation, the $\mu\text{p}(F=1)$ atom
  quickly de-excites back to the $F=0$ state in collisions with protons
  bound in the surrounding H$_2$ molecules.
  As a result, the HFS energy
  is converted into additional $\mu$p kinetic energy up to~0.12~eV
  with a wide distribution due the simultaneous
  rotational transitions in the affected H$_2$ molecules.

  A clear and strong signal from the ``laser-ac\-ce\-le\-ra\-ted'' neutral
  $\mu$p atoms is necessary in order to perform a precise HFS
  measurement.
  Our idea is to use an admixture of a high-Z element
  in the H$_2$ target and detect characteristic X-rays from the
  de-excitation of the high-Z muonic atom formed when the muon
  is transferred from $\mu$p. Since muon transfer
  takes place both from the thermalized and laser-ac\-ce\-le\-rated $\mu$p
  atoms, it is crucial to choose a higher-Z element which is
  characterized by a strong energy dependence of the 
  muon-transfer rate below about 0.1~eV.

  According to the PSI experiments~\cite{schneuvly88,werthmuller96,werthmuller98},
  oxygen is a good
  candidate. Although these experiments were performed only at room
  temperature, an unexpectedly high value of the muon transfer rate was
  found for the ($\mu$p)$_{1S}$
  atoms not yet thermalized (``epithermal''). 
  Energetic $\mu$p atoms are 
  due to the acceleration via atomic cascade~\cite{jensen02,faifman07}
  after the negative muon capture in hydrogen and $\mu$p formation in
  excited Coulomb states ($n\approx{}14$). A strong dependence of the
  muon transfer rate to oxygen to the collision energy was also
  anticipated in the quantum-mechanical calculations reported in Ref.~\cite{dupays04,lelin05}.

  The PSI data on the energy-dependent muon
  transfer rate to oxygen
  were fitted with a two-step function of the collision
  energy~\cite{werthmuller98}. 
  This function, however, is inappropriate for modelling
  the HFS measurement because (a)
  the fit is too rough an approximation to the data
  and (b), most
  important, it provides no information about the energy
  dependence of the transfer rate in the energy range below
  the threshold of 0.12 eV (corresponding to
  temperatures $\sim$ 1000 K), which in fact is the range of
  experimental interest. This information can only be obtained by
  measurements of the muon transfer rate to oxygen from $\mu$p 
  thermalized atoms in a sufficiently wide temperature range
  below 1000~K, as performed in the present work.
  
  In thermal equilibrium at temperature $T$, the observable transfer rate
  $\Lambda_{\rm pO}(T)$ is expressed in terms of the function that
  describes the dependence of the transfer rate on the collision
  energy $E$ by means of the relation
  \begin{equation}
    \Lambda_{\text{pO}}(T)=\int\limits_0^{\infty}dE\,f_{\rm MB}(E;T)\,
    \lambda_{\text{pO}}(E),
    \label{eq:T2Egen}
  \end{equation}
  where 
  \begin{equation}
    f_{\rm MB}(E;T)=2\sqrt{E/\pi} \,(k_B T)^{-3/2}\,e^{-E/(k_B T)}
  \end{equation}
  is the Max\-well--Bol\-tzman distribution, with $k_B$ the Boltzman
  constant. 
  The main contribution to the integral in the right-hand side of
  Eq.~(\ref{eq:T2Egen}) comes from the region around
  $E=k_B T$; therefore the measurement of $\Lambda_{\text{pO}}(T)$ in
  the range between 100 and 300~K 
  will provide important information about the behavior of
  $\lambda_{\text{pO}}(E)$ for energies in and around the range
  $0.01<E<0.05$~eV.

  The obtained results allow us to test the available theoretical calculations of the
  muon transfer rate and will be used
  to choose optimal conditions for the FAMU HFS measurement.

\section{Experiment}
  The FAMU apparatus, used for the transfer rate measurement, is built
  around a pressurized and thermalized gas target contained in
  an aluminium vessel kept at thermal equilibrium and surrounded by
  X-ray detectors.
  The FAMU experimental method requires a detection system
  suited for time resolved X-ray spectroscopy~\cite{adamczak16}.
  Eight scintillating counters are used, based on
  1'' Lanthanum Bromide crystals and read by Hamamatsu high speed
  photomultipliers. Data are acquired and processed by 500~MHz CAEN
  V1730 digitizers, within the framework of the FAMU Data Acquisition
  System~\cite{soldani}. Output signals from the detectors provide
  measurements of both the energy and time spectrum of the recorded
  events. In this configuration, the detectors provide
  a good resolution, about 8.5\% (FWHM) at 133~keV, and an excellent
  time resolution, signal rise time of about 12~ns~\cite{famu2016paper}. 
  The experiment was performed at the RIKEN muon facility of
  the Rutherford Appleton Laboratory (UK)~\cite{RAL}, using muons produced in bunches with a repetition rate of 50~Hz.
  More details about
  the FAMU apparatus can be found in~\cite{famu2016paper}.

  The data used in this analysis were acquired using
  a gas mixture of H$_2$ and O$_2$ with an oxygen concentration
  $c_{\text{O}}$ of 190~ppm. The target was
  filled at room temperature to 41~bar and subsequently brought to six different
  temperatures, from 300 to 104~K degrees. The lowest temperature was
  significantly higher than the oxygen condensation point for the
  mixture, 54~K, obtained from Ref.~\cite{stewart}.
  A thermally insulated vertical tube provided external connection to the target. After filling, the target was 
  sealed by a valve placed just outside the cryogenic vessel. The portion 
  of tube between the target and the valve had a volume about ten times 
  smaller than the target itself. Due to gravity, the temperature 
  gradient inside the tube ranged from room temperature, at the valve, to 
  the gas target temperature, at the vessel. The 
  variation of density inside the target was sufficiently small that any associated systematic effects were negligible. 

  An accurate determination of the muon transfer rate as a function of
  kinetic energy requires measurements under steady-state conditions, when
  the kinetic-energy distribution of $\mu$p atoms is well known. 
  The~distribution of $\mu$p initial energy, after the muon capture in
  H$_2$ and subsequent atomic cascade, is broad and can cover hundreds of ~eV
  depending on the target density~\cite{jensen02,faifman07,pohl06,bader}.
  The shape of this distribution is not well known. Therefore,
  we have measured the muon transfer rate only after the slowing down and
  thermalization of $\mu$p atoms, when their kinetic energies are
  described by the Maxwell--Boltzmann distribution for a given target
  temperature.

  Data were taken with a target number density which enabled
  fast thermalization of $\mu$p atoms (many times faster than the muon
  lifetime) and rapid quenching of the initial statistical population
  of the spin state $F=1$. For H$_2$ gas at 41~bar and 300~K, the Monte
  Carlo simulation gives a thermalization time of 150~ns and a
  quenching time of 10~ns~\cite{bakalov15b}. In order to observe the
  characteristic X-rays from muonic oxygen over long times, the oxygen
  concentration for a given target density cannot be too high. For
  $c_{\text{O}}=190$~ppm at this temperature and pressure, the average muon
  transfer rate from the thermalized $\mu$p atoms to oxygen is
  $0.78\times{}10^6$~s$^{-1}$ (using the experimental rate of
  $8.5\times{}10^{10}$~s$^{-1}$ at liquid-hydrogen %
  density~\cite{werthmuller98}), which is comparable with the muon
  decay rate. This choice enabled us to observe the muon-transfer
  process from thermalized $\mu$p atoms for several microseconds.

  Muonic oxygen atoms are formed in excited Coulomb states, which promptly de-excite by emitting characteristic X-rays. A measurement
  of the muon transfer rate is performed by studying the time evolution
  of muonic-oxygen lines. The muon-transfer process was studied at
  times beyond the end of the muon pulse and
  much larger than the prompt emission and $\mu$p thermalization time.
  In this way the large background of X-ray emissions due to the
  interaction between muons and all the elements of the target ---
  mostly aluminium, nickel, gold, and carbon --- was strongly suppressed
  and could be neglected.

  Under these conditions, at a given temperature, $T$, the variation of the number $N_{\mu\text{p}}$ of $\mu{}$p atoms
  in the target in the time interval $dt$ is given by:  
  \begin{equation}
    dN_{\mu\text{p}}(t) = - N_{\mu\text{p}}(t)\lambda_{\text{dis}}(T)\, dt \,,
    \label{eq:tevol}
  \end{equation}  
  where $\lambda_{\text{dis}}(T)$ is the total disappearance
  rate of muonic hydrogen atoms at temperature $T$, given by:
  \begin{equation}
    \lambda_{\text{dis}}(T) = \lambda_0+\phi\,[c_{\text{p}}\Lambda_{\text{pp}\mu}+
    c_{\text{d}}\Lambda_{\text{pd}}(T) +c_{\text{O}}\Lambda_{\text{pO}}(T) ].
    \label{eq:ldis}
  \end{equation}
  Here $\lambda_0 = (4665.01\pm0.14)\times 10^2$~s$^{-1}$~\cite{andreev,suzuki} is 
  the disappearance rate of the muons bound to
  protons (that includes both muon decay and nuclear capture),
  $\Lambda_{\text{pp}\mu}=2.01\times 10^6$~s$^{-1}$~\cite{andreev} is the formation
  rate of the $pp\mu$  molecular ion in
  collisions of $\mu$p with a hydrogen nucleus (normalized to liquid
  hydrogen density, LHD, $N_0=4.25\times 10^{22}$~atom/cm$^{3}$) and $\Lambda_{\text{pO}}(T)$ is the muon
  transfer rate from $\mu$p to oxygen atoms. 
  The muon transfer rate $\Lambda_{\text{pd}}(T)$ from $\mu$p to
  deuterium, bound mostly in HD molecules, varies from
  $8.65\times{}10^{9}$~s$^{-1}$ at 100~K to
  $8.20\times{}10^{9}$~s$^{-1}$ at 300~K (normalized to $2.125\times 10^{22}$~HD~molecules/cm$^{3}$). The rate
  $\Lambda_{\text{pd}}$ for muon transfer to bare deuterium nuclei is
  practically constant at the lowest energies~\cite{chiccoli}. The
  energy-dependent electron screening in hydrogenic
  molecules~\cite{adamczak06} leads to the above appreciable change of 
  $\Lambda_{\text{pd}}(T)$ in the considered temperature interval. The number density of the atoms in the target gas is $\phi=
  (4.869\pm0.003)\times10^{-2}$ in LHD units, and
  $c_{\text{p}}$, $c_{\text{O}}$, and $c_{\text{d}}$ are the number concentrations of
  hydrogen, oxygen, and deuterium respectively. We used hydrogen
  with measured natural deuterium abundance
  $c_{\text{d}}=(1.358\pm0.001)\times
  10^{-4}$~\cite{boschi}\footnote{A sample
    bottle of the same batch of pure hydrogen used to produce the
    mixture was provided by the gas supplier. This sample gas was tested using a precise mass
    spectrometer to determine the deuterium abundance. Unfortunately,
    the same study could not be done with the gas mixture used during
    the acquisition to determine better the oxygen abundance.}. The oxygen
  concentration was $c_{\text{O}} = (1.90\pm0.04)\times 10^{-4}$ and
  $c_{\text{p}}=1-c_{\text{O}}-c_{\text{d}}$.
  
  Let us note that the rate of nonresonant $pp\mu$ formation is
  practically constant below 0.1~eV~\cite{faifman89}. 
  The $\mu$d atoms, which are created via the muon transfer from $\mu$p 
  atoms to the small natural admixture of deuterium, have an initial
  energy of about 42~eV. They are never thermalized since a deep
  Ramsauer-Townsend minimum in the cross section of scattering
  $\mu$d+H$_2$ is present at 7~eV~\cite{adamczak96}. As a result, for
  target densities corresponding to 300~K hydrogen at 41~bar, the 
  mean kinetic energy of $\mu$d atoms is 26~eV and only a very
  small fraction of these atoms is thermalized. The muon transfer rate
  $\Lambda_\text{dO}$ from $\mu$d atoms to oxygen is negligible at
  energies $\gtrsim{}1$~eV, as confirmed by the absence of
  corresponding X-ray spectra from muonic oxygen at short times in
  the experiment performed in a pure D$_2$ target with a small
  admixture of SO$_2$~\cite{mulhauser93}. Therefore, a contribution
  to the X-ray spectra due to the muon transfer from $\mu$d atoms to
  oxygen is neglected in the present analysis. Our Monte Carlo
  simulation, which used the experimental results of %
  Ref.~\cite{mulhauser93} for the rate $\Lambda_\text{dO}$ at thermal,
  epithermal and higher energies, has shown that this contribution is
  smaller than 1\% for times up to several microseconds.

\section{Data analysis}
  In our analysis, only the steady-state delayed X-ray events due to the thermalized
  atoms are considered. The only unknown variable is the transfer rate
  $\Lambda_{\text{pO}}(T)$, which is determined using Eqs.~(\ref{eq:tevol}) and
  (\ref{eq:ldis}) by numerically fitting the time evolution of the oxygen X-rays  at temperature $T$ and leaving $\Lambda_{\text{pO}}$ a free parameter.
  Details about the method can be found in Ref.~\cite{mocchiutti18b}.

  The experimental sample used in this work consists of about
  $2.6\times 10^6$~muon triggers, and corresponds to
  $\approx 7.8\times 10^7$ reconstructed X-rays. The time, in nanoseconds, associated
  to each X-ray is relative to the trigger generated by the accelerator and beam control system. In this time reference
  frame, the two muon spill bunches peak at about 530 and 850 ns. Each
  muon spill bunch consists of about $10^3$ muons.

  Data were taken at six target temperatures: 104, 153, 201, 240, 272, and
  300~K. Two sensors were used to measure the target
  temperature. The cryogenic system was able to keep the temperature
  stable by limiting fluctuations to about 10~mK/h, within the
  systematic errors of the temperature sensors. More details about
  the cryogenic system can be found in~\cite{famu2016paper}. 
  Each temperature was kept stable for an acquisition
  time of three hours. 

  X-ray signals were identified and reconstructed using a fitting
  procedure on the detector waveforms. A clean data sample was
  obtained by applying light selection criteria based on the reduced
  $\chi^2$ of the waveform fit and on the distance between two
  consecutive reconstructed signals. 
  The reduced $\chi^2$ selection ($\tilde{\chi}^2 < 100$) was used to reject
  not-converged fit and poorly reconstructed events. The requirement on
  the distance between two consecutive signals to be greater than 30~ns
  was chosen by testing the reconstruction algorithm by means of a
  GEANT4 simulation. The simulation showed that when two peaks are
  separated by more than 30~ns the software reconstruction efficiency
  and accuracy (correct energy reconstruction) is better than 99.9\%. 
  Selection efficiencies and
  fractional live time were estimated and taken into account in the analysis of the time evolution of the oxygen lines. The combined efficiencies and
  fractional live time were about 95\%, constant above 2000 ns, and
  smoothly decreased to about 92\% at 1200 ns.
  A detailed discussion of the data
  selection and selection efficiencies can be found in Ref.~\cite{mocchiutti18b}.
  For each temperature, the
  energy spectrum of delayed events was studied as a function of time.
  The delayed time window -- from $\approx$1200 to 10000~ns from the trigger -- was split
  into 20~bins of increasing width, the narrowest being $\approx 140$~ns. The
  time resolution of the reconstructed events was better than 1~ns,
  hence any migration effect on neighbouring bins was negligible and no time
  deconvolution was needed.

  An estimation of the background below the oxygen-line signal was the
  most important aspect of the data analysis. The
  background was estimated for each time and temperature bin using the
  data taken with a pure H$_2$-gas target and with the same selection
  criteria and same gas condition, but with a smaller statistical sampling.

  \begin{figure*}[!thb]
    \includegraphics[width=0.95\textwidth]{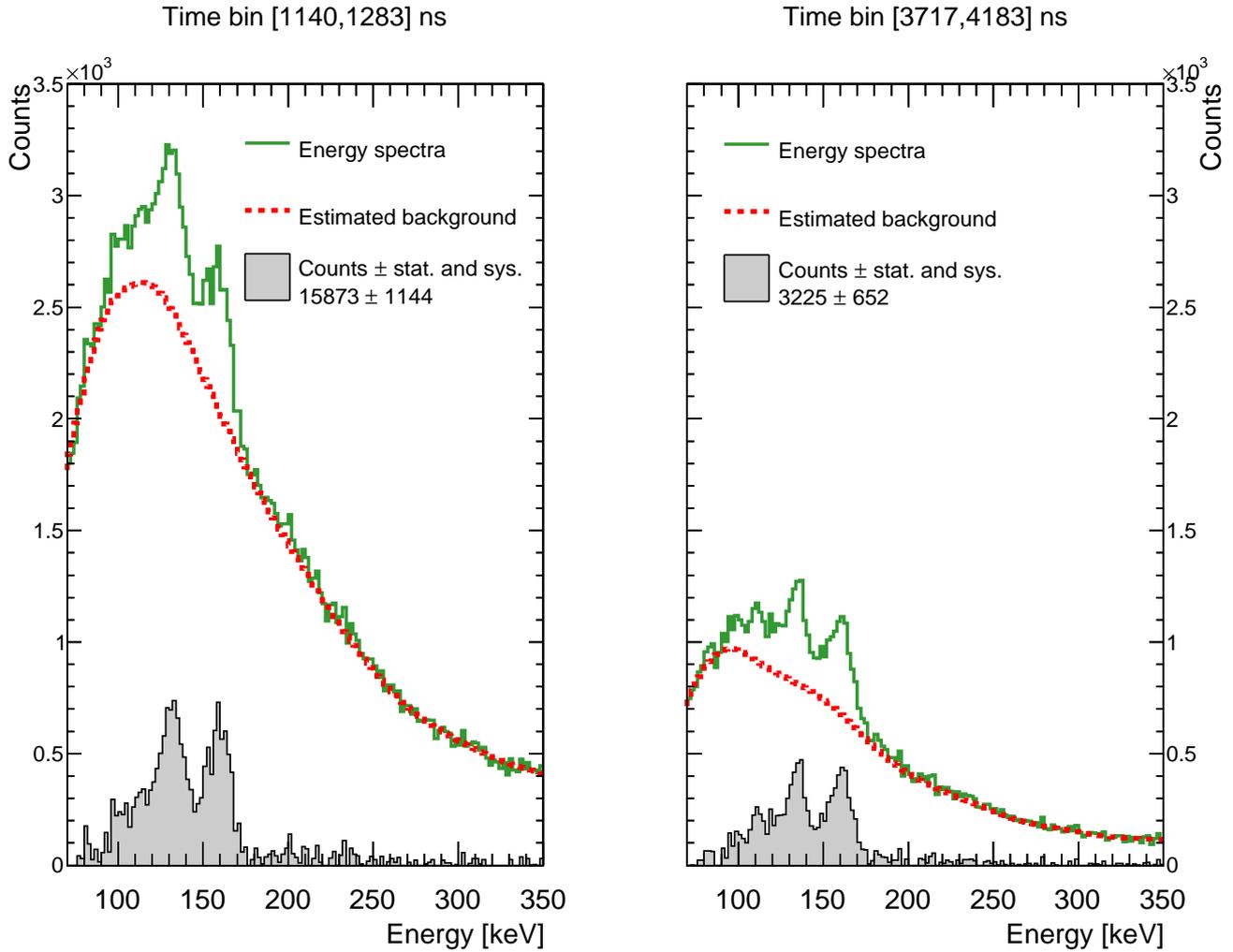}\hspace{2pc}
    \caption{Energy spectra at 104~K in two different time bins: [1140,1283]~ns
      in the left panel and [3717,4183]~ns in the right panel. Solid
      green line: energy spectra; dotted red line: estimated
      background; shaded area: signal after background subtraction.\label{fig1}}
  \end{figure*}
  Figure~\ref{fig1} shows the energy spectra at 104~K in two different
  time bins. These energy spectra were obtained by summing the
  calibrated energy spectra of each single LaBr detector. 
  Dotted lines represent the background (mostly induced by
  electrons from muon decay) estimated using a target of pure hydrogen
  gas. Due to the lower statistics, the fluctuations in the background
  spectra were higher, hence, the spectra were smoothed using a Gaussian kernel
  algorithm~\cite{shapiro}. The background was normalized to the data,
  for each time bin and energy spectrum, in the region between 250 and 350~keV. 
  In the figure, the signal after background subtraction is drawn as a shaded
  area. The tails towards lower energy are due to energy leakage from
  the LaBr crystal. This effect was studied by means of GEANT4
  simulations and corresponds to X-rays hitting the crystals on the
  border, with an incoming direction almost parallel to the surface.
  The total number of detected oxygen
  X-rays in the range 100~to 200~keV varied from about ten thousands to few hundreds, depending
  on target temperature and time bin.
  Systematic errors due to the normalization of the background were
  estimated by using a statistical approach on the
  number of both background and signal events in the normalization
  interval. A further consideration of systematic errors in the shape of the background was
  estimated by comparing the pure hydrogen background distribution
  with analytical functions fitted to the data and by using gas target
  mixture with different composition (i.e., CO$_2$ in hydrogen and Ar
  in hydrogen).
  The overall systematic error was quadratically summed to the statistical error for each signal
  spectrum (see, e.g., numbers reported in Fig.~\ref{fig1}), before performing the fit to the data that provides the
  transfer-rate measurement.

  Notice that, due to the background subtraction, a time measurement
  cannot be associated to each oxygen x-ray, i.e. it is not possible
  to tag the single event. 
  Hence, after
  background subtraction in a given time bin, there is not an average time measurement
  coupled to the signal spectrum. 

  \begin{figure*}[!thb]
    \includegraphics[width=0.95\textwidth]{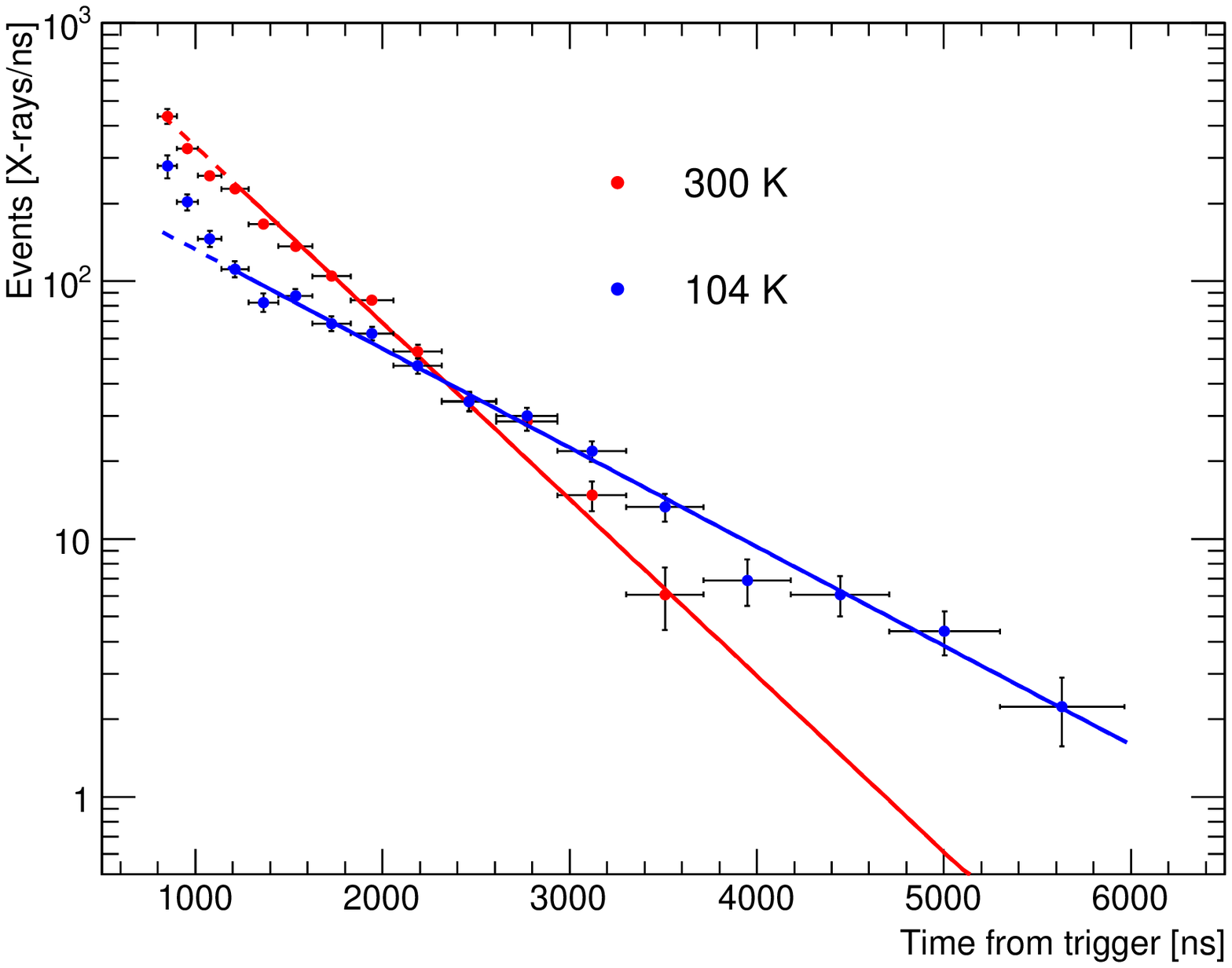}\hspace{2pc}
    \caption{Time dependence of oxygen line intensities at 300~K and 104~K. Two
      exponential functions (solid
      lines) correspond to the transfer rate fit. Dashed lines are an
      extrapolation of the fit at shorter times. Vertical error bars associated to the points include
      the statistical and background-systematic errors. Horizontal
      error bars represent the bin width.\label{fig2}}
  \end{figure*}
  The time evolution of the oxygen line X-ray spectra, at the two temperature
  limits of 300~K (red circles) and at 104~K (blue circles), is shown in
  Fig.~\ref{fig2} and reported in Table~\ref{tabdata} for all the
  temperatures. 
  \begin{table*}[!ht]    
    \begin{center}
      \caption{
      Time evolution of the oxygen line X-ray spectra measured at six
      different temperatures. Uncertainties associated to the
      measurements include statistical and background-systematic
      errors. Notice that the first three time bins cover a
      non-thermalized phase which coincide with the second muon beam spill
      arrival, low detection efficiency, and high pile-up effects. The
      fit was performed starting from the bin $[1140,1283]$~ns.      
      \label{tabdata}}
      \vspace{0.5pt}
      \begin{tabular}{c|c|c|c|c|c|c}
        \hline
        \hline
        Time bin &  \multicolumn{6}{c}{Oxygen line time evolution [X-rays/ns]}  \\
        \cline{2-7}
          $[$ns$]$   &  300~K & 272~K & 240~K & 201~K & 153~K & 104~K \\
        \hline
          800 -- 900 & 435 $\pm$ 29 & 477 $\pm$ 28 & 451 $\pm$ 29 & 407 $\pm$ 28 & 365 $\pm$ 30 & 279 $\pm$ 28 \\ 
         900 -- 1013 & 326 $\pm$ 15 & 333 $\pm$ 14 & 296 $\pm$ 15 & 266 $\pm$ 14 & 220 $\pm$ 15 & 202 $\pm$ 14 \\ 
        1013 -- 1140 & 257 $\pm$ 11 & 252 $\pm$ 10 & 243 $\pm$ 11 & 224 $\pm$ 10 & 182 $\pm$ 11 & 146 $\pm$ 10 \\ 
        1140 -- 1283 & 228 $\pm$  8 & 216 $\pm$  8 & 215 $\pm$  8 & 192 $\pm$  8 & 156 $\pm$  8 & 111 $\pm$  8 \\ 
        1283 -- 1444 & 166 $\pm$  7 & 165 $\pm$  7 & 161 $\pm$  7 & 133 $\pm$  7 & 123 $\pm$  7 &  83 $\pm$  7 \\ 
        1444 -- 1626 & 136 $\pm$  5 & 124 $\pm$  5 & 133 $\pm$  5 & 119 $\pm$  5 & 109 $\pm$  6 &  88 $\pm$  5 \\ 
        1626 -- 1829 & 104 $\pm$  5 & 104 $\pm$  4 & 103 $\pm$  5 &  96 $\pm$  4 &  87 $\pm$  5 &  68 $\pm$  4 \\ 
        1829 -- 2059 &  84 $\pm$  4 &  85 $\pm$  4 &  90 $\pm$  4 &  83 $\pm$  4 &  76 $\pm$  4 &  63 $\pm$  4 \\ 
        2059 -- 2317 &  53 $\pm$  3 &  60 $\pm$  3 &  58 $\pm$  3 &  57 $\pm$  3 &  55 $\pm$  3 &  47 $\pm$  3 \\ 
        2317 -- 2608 &  34 $\pm$  3 &  36 $\pm$  3 &  36 $\pm$  3 &  40 $\pm$  3 &  37 $\pm$  3 &  34 $\pm$  3 \\ 
        2608 -- 2935 &  28 $\pm$  2 &  26 $\pm$  2 &  26 $\pm$  2 &  30 $\pm$  2 &  31 $\pm$  2 &  30 $\pm$  2 \\ 
        2935 -- 3303 & 14.8$\pm$1.9 & 15.1$\pm$1.9 &  14 $\pm$  2 & 17.1$\pm$1.9 &  20 $\pm$  2 & 21.9$\pm$1.9 \\ 
        3303 -- 3717 &  6.1$\pm$1.6 &  7.1$\pm$1.6 & 9.9 $\pm$1.6 & 10.2$\pm$1.6 & 11.9$\pm$1.7 & 13.3$\pm$1.6 \\ 
        3717 -- 4183 &          --- &          --- & 4.2 $\pm$1.3 & 4.6 $\pm$1.3 & 6.2 $\pm$1.4 & 6.9 $\pm$1.4 \\ 
        4183 -- 4708 &          --- &          --- &          --- & 4.2 $\pm$1.0 & 4.5 $\pm$1.1 & 6.1 $\pm$1.1 \\ 
        4708 -- 5298 &          --- &          --- &          --- &          --- &          --- & 4.4 $\pm$0.8 \\ 
        5298 -- 5963 &          --- &          --- &          --- &          --- &          --- & 2.2 $\pm$0.7 \\ 
        \hline
        \hline
      \end{tabular}  
    \end{center}
  \end{table*}
  Each point represents the integrated signal, after the
  background subtraction and efficiency correction, divided by the
  time bin width. Data point are centered in the time bins and error
  bars on the x axis represent the time bin width. 
  To perform a fit of the data versus
  time, a step-like fit function was used: the fit function is an
  exponential function, which is integrated within each time bin
  that was defined for the collected data. Function and data can then be considered as two histograms,
  which can be compared directly. For the purposes of this
  quantitative comparison the question of choosing the correct time
  value within the bin is irrelevant. 
  The fit was performed starting at $\approx$1200~ns, about 350~ns after the
  second muon spill in order to take into account only the thermalized
  phase, given a thermalization time of 150~ns. 
  The upper limit of the
  time window was
  chosen according to the available statistics: data points were used
  if the measured integrated signal was greater than three times the
  associated error. 
  Figure~\ref{fig2} shows that the slopes of the two
  distributions differ 
  significantly, corresponding to the different
  transfer rates as functions of temperature. 
  Solid lines represent the exponential transfer rate obtained by the fit. The
  comparison between data points and curves has to be done considering
  the bin width (error bars on the x axis). Dashed line 
  lines are the extrapolation of the fit to shorter times. The first
  three data points were not included in the fit since they are too
  close to the muonic hydrogen production, hence the $\mu$p are still
  energetic and in an epithermal state. The same break in the spectra
  was observed at PSI and was helpful in understanding the existence
  of a kinetic energy dependence of the transfer rate from muonic
  hydrogen to oxygen~\cite{werthmuller98}. The error bars associated to
  the points include the statistical and background-systematic errors,
  as described previously.

\section{Results}
  The experimental results are presented in Table~\ref{tab}. During the 
  measurements the temperature was kept stable, with a variation around
  the mean value $T$ less than 0.1\%
  (column~1), to which the measured rate $\Lambda_{\text{pO}}$ (column~2) is referred.
  \begin{table*}[!ht]    
    \begin{center}
      \caption{Summary of transfer rates from muonic hydrogen to oxygen. The first error represents the statistical and background related systematic errors quadratically summed. The second error reports general systematic uncertainty. The third column reports the reduced $\chi^2$ of the
        fit. The fourth and fifth columns show the results of the
        Kolmogorov-Smirnov (K-S) test and its derived maximum distance, respectively,
        performed on the pulls in comparison with a standard Gaussian distribution.\label{tab}}
      \vspace{0.5pt}
      \begin{tabular}{cccccc}
        \hline
        \hline
        Mean & $\Lambda_{\text{pO}}(T)$ & Reduced & K-S test & K-S max\\
        temperature $T$ [K] & [$10^{10} s^{-1}$] & $\chi^2$ & on pull & distance\\ \hline 
        104 & $  3.07 \pm   0.29 \pm    0.07$ & 1.21 & 1.00 & 0.14 \\  
        153 & $  5.20 \pm   0.33 \pm    0.10$ & 0.81 & 0.85 & 0.25 \\ 
        201 & $  6.48 \pm   0.32 \pm    0.13$ & 1.95 & 1.00 & 0.17 \\ 
        240 & $  8.03 \pm   0.35 \pm    0.16$ & 1.64 & 0.81 & 0.27 \\ 
        272 & $  8.18 \pm   0.37 \pm    0.17$ & 1.69 & 0.76 & 0.30 \\ 
        300 & $  8.79 \pm   0.39 \pm    0.18$ & 1.80 & 0.76 & 0.30 \\ 
        \hline
        \hline
      \end{tabular}  
    \end{center}
  \end{table*}
  For each fit, the reduced $\chi^2$ is reported (third column). The
  quality of the fit was assessed also by studying the pull
  distribution, where the pull is defined as the ratio between the
  residual and the error associated to the point ($(X_{\mbox{measured}} - X_{\mbox{fitted}}) / \sigma_X$). If the fit is good and the fluctuation
  purely statistical, then the pulls are distributed as a
  standard normal distribution (a Gaussian with zero mean and unit width). For  each fit, the unbinned and ordered pulls data set was compared to a  standard normal distribution using the Kolomogorov-Smirnov test. Results of
  probability and maximum distance are reported in columns four and
  five of Table~\ref{tab}. The results of the Kolmogorov-Smirnov test applied to the six
  measurements could not exclude normally distributed pulls.

  \begin{figure*}[!thb]
    \includegraphics[width=0.47\textwidth]{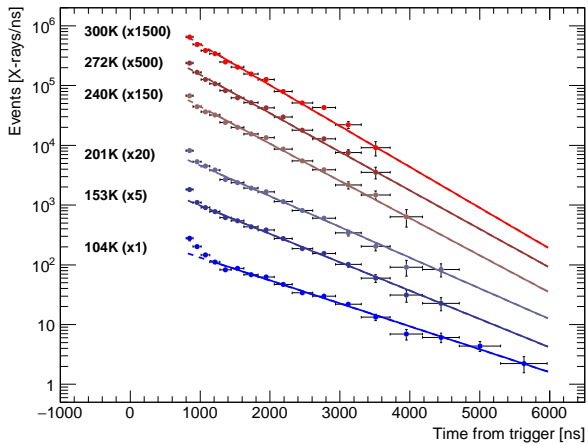}\hspace{2pc}
    \includegraphics[width=0.47\textwidth]{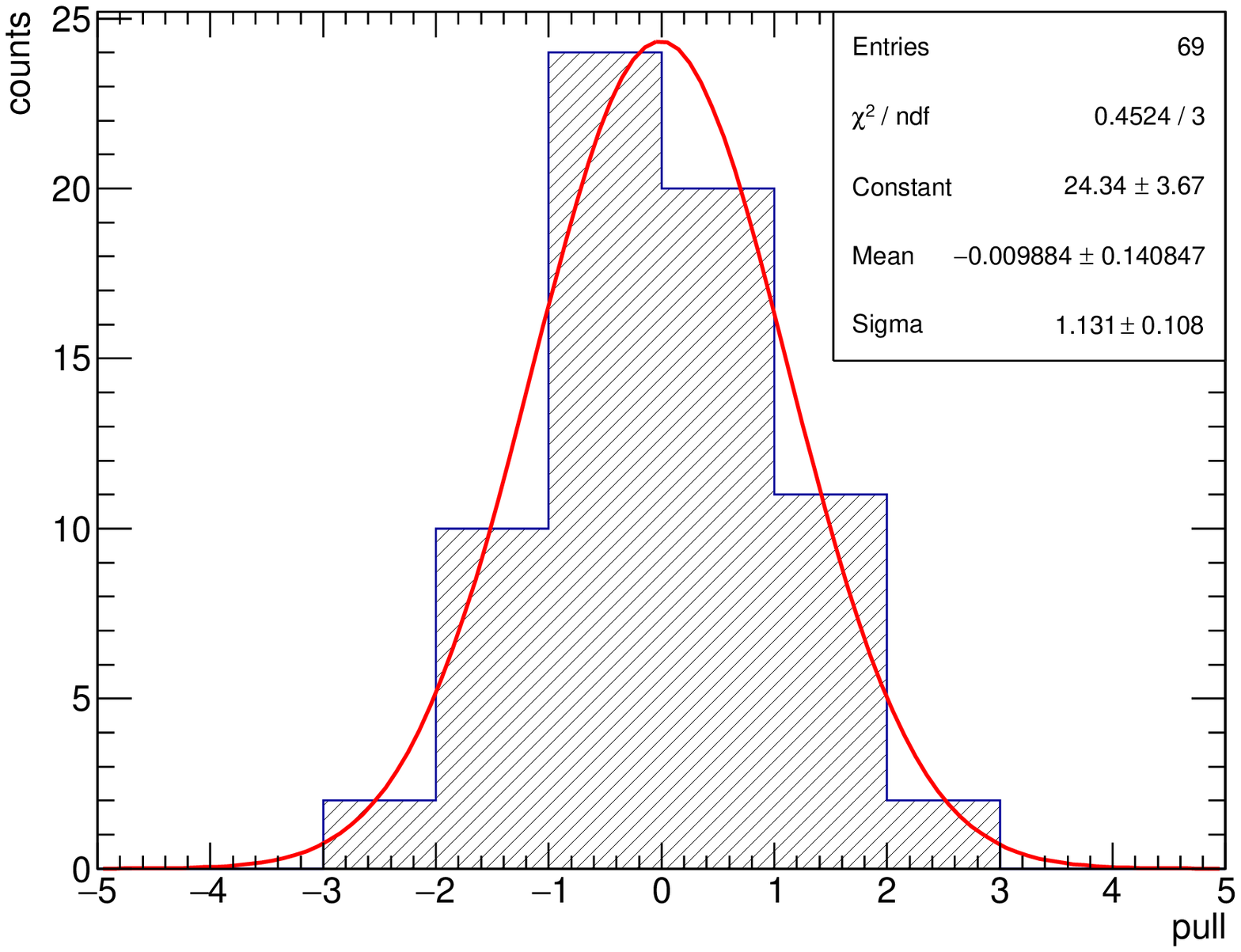}\hspace{2pc}
    \caption{Left panel: time dependence of the oxygen line
      intensity for the six temperature bins. Each set of data has
      been scaled by the factor reported in the figure. Error bars, solid and
      dashed lines have the same meaning as in Fig.~\ref{fig2}. Right
      panel: overall pull distribution. Solid line is a Gaussian fit of the
      data. Mean and width are consistent with a standard normal distribution. \label{fig3}}
  \end{figure*}
  The six data sets and corresponding fits are reported also in
  Fig.~\ref{fig3}, left panel, where each one is artificially adjusted vertically for clearer visualization.  It can be seen that the epithermal component 
  becomes more evident at low temperatures when the thermal transfer
  rate is smaller. The right panel of Figure~\ref{fig3} shows the overall distribution of the pulls for the  six temperature fits. The solid line represents a Gaussian fit to
  the distribution and it is in excellent agreement with a standard normal distribution.

  \begin{figure*}[!thb]
    \includegraphics[width=0.95\textwidth]{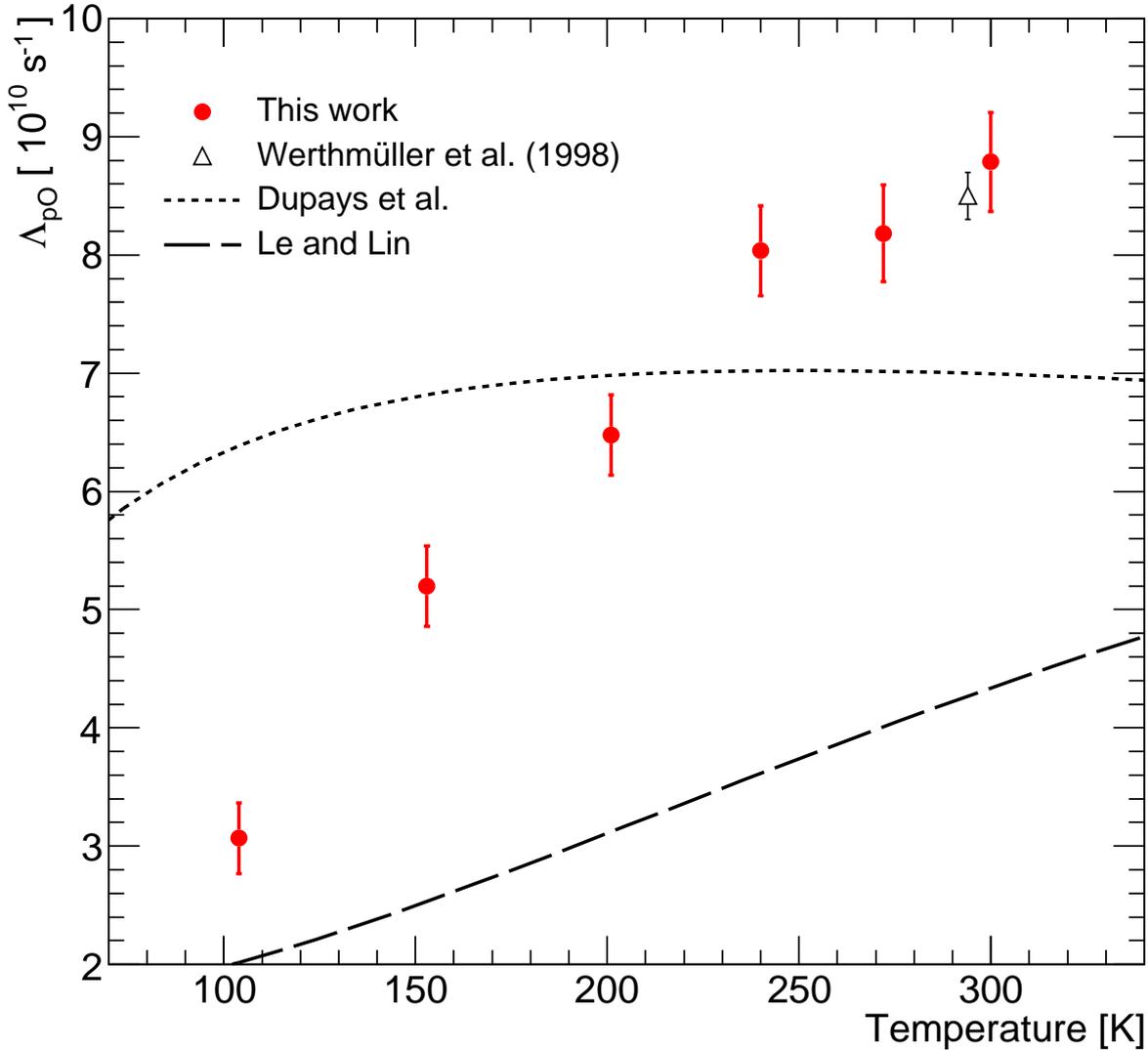}\hspace{2pc}
    \caption{Transfer rate from muonic hydrogen to oxygen:
    comparison of the present work with the
    experimental~\cite{werthmuller98}, and the
    theoretical results~\cite{dupays04,lelin05} converted in their
    temperature dependence using Maxwell--Boltzmann distributions.\label{fig4}}
  \end{figure*}
  Figure~\ref{fig4} shows the results obtained in this work. This is the first measurement of the transfer rate
  from muonic hydrogen to oxygen as function of the temperature. 
  The error bars represent the quadratic sum of statistical and
  systematic errors, as reported in Table~\ref{tab}. 
  The analysis of systematic errors
  shows that the main source of uncertainty derives from the gas
  composition. The H$_2$/O$_2$ gas mixture was prepared by the supplier with a
  relative uncertainty of 3\%. Other sources of systematic effects, including
  temperature and pressure measurements, timing, parameters error
  propagation were estimated to be smaller than 1\% each and considered negligible.
  Results are in excellent  agreement with the PSI measurement at
  294~K~\cite{werthmuller98}. The lines represent the theoretical
  results presented in~\cite{dupays04,lelin05} as function of kinetic energy, 
  which we converted to temperature using~Eq.~\ref{eq:T2Egen}. 
  Experimental data are not in agreement with
 theoretical calculations, however. This is not surprising, since these rough
  calculations do not take into account in details the electron
  screening effects and also other aspects are simplified, e.g. oxygen is treated as a
  free atom and not as part of oxygen molecule.

\section{Conclusions}
  In the first investigation of the
  temperature dependence of the
  muon-transfer process from the thermalized $\mu$p atoms to oxygen,
  we have observed a strong monotonic rise by a factor of $\sim3$ of the rate
  $\Lambda_{\text{pO}}(T)$ in the temperature interval 104--300~K.
 
  The measurements were performed in conditions of thermal
  equilibrium that allows us to use Eq.~\ref{eq:T2Egen} and
  anticipate a much greater rise 
  of the muon transfer rate
  $\lambda_{\rm pO}(E)$ from energies $E\sim0.01$ eV up to
  energies of the order of $E\sim 0.1$ eV. The work on extracting the explicit energy dependence and 
  the uncertainty of $\lambda_{\rm pO}(E)$ from the experimental data in
  Table~\ref{tab} is currently in progress; the results will be
  presented elsewhere.
  
  Such a strong change enables us to employ the muon
  transfer rate to oxygen as a signature of the kinetic-energy gain of
  the $\mu$p atom in the planned FAMU spectroscopy measurement of the hyperfine
  splitting of 1$S$ state of this atom.

  These results allow to test the available theoretical methods for the
  calculation of charge transfer processes in non-elastic atom scattering
  and will hopefully stimulate the development of new and more
  efficient computational approaches.

\section*{Acknowledgments}
  The research activity presented in this paper has been carried out in
  the framework of the FAMU experiment funded by Istituto Nazionale di
  Fisica Nucleare (INFN). The use of the low energy muons beam has been
  allowed by the RIKEN RAL Muon Facility. We thank the RAL staff
  (cooling, gas, and radioactive sources sections) and especially Mr.
  Chris Goodway, Pressure and Furnace Section Leader, for their help,
  suggestions, professionalism and precious collaboration in the set-up
  of the experiment at RIKEN-RAL.
  
  We thank the Criotec company and especially Ing.~Adri\-a\-no
  Mussinatto for the technical help and support in the construction of
  the FAMU target.
  
  We thank our colleagues Chiara Boschi and Ilaria Ba\-ne\-schi (IGG,
  CNR Pisa) for their help in the measurement of the gas isotopic
  composition.
  
  D.~Bakalov, P.~Danev and M.~Stoilov acknowledge the support of Grant
  DN08-17 of the Bulgarian Science Fund.
  
\bibliographystyle{elsarticle-num}
\bibliography{famu2016_pla_v4}

\end{document}